\documentstyle[emulateapj]{article}

\newcommand{\kms}{km~s$^{-1}$}

\input epsf

\newcommand{\lya}{Ly$\alpha$}
\newcommand{\lyb}{Ly$\beta$}
\newcommand{\lyg}{Ly$\gamma$}
\newcommand{\lyd}{Ly$\delta$}
\newcommand{\lye}{Ly$\epsilon$}

\begin{document}

\title{FUSE OBSERVATIONS OF THE LOW-REDSHIFT \lyb\ FOREST }
\author{ J. MICHAEL SHULL\altaffilmark{1,2},  
MARK L. GIROUX\altaffilmark{1},
STEVEN V. PENTON\altaffilmark{1},
JASON TUMLINSON\altaffilmark{1},
JOHN T. STOCKE\altaffilmark{1},
EDWARD B. JENKINS\altaffilmark{3},
H. WARREN MOOS\altaffilmark{4},
WILLIAM R. OEGERLE\altaffilmark{4},
BLAIR D. SAVAGE\altaffilmark{5},
KENNETH R. SEMBACH\altaffilmark{4},
DONALD G. YORK\altaffilmark{6},
JAMES C. GREEN\altaffilmark{1},
and BRUCE E. WOODGATE\altaffilmark{7}
}
\altaffiltext{1}{CASA and Dept. of Astrophysical and Planetary Sciences,
   University of Colorado, Boulder, CO 80309}
\altaffiltext{2}{Also at JILA, University of Colorado and National 
   Institute of Standards and Technology}
\altaffiltext{3}{Dept. of Astrophysical Sciences, Princeton University,
  Princeton, NJ 08544}
\altaffiltext{4}{Dept. of Physics \& Astronomy, Johns Hopkins University, 
   Baltimore, MD 21218}
\altaffiltext{5}{Dept. of Astronomy, University of Wisconsin, Madison,
   WI 53706}
\altaffiltext{6}{Dept. of Astronomy \& Astrophysics, Univ. of Chicago, 
   Chicago, IL 60637}  
\altaffiltext{7}{NASA Goddard Space Flight Center, Greenbelt, MD 20771}

\begin{abstract}

We describe a moderate-resolution (20--25 \kms) FUSE study of the 
low-redshift intergalactic medium.  We report on studies of
7 extragalactic sightlines and 12 \lyb\ absorbers that correspond to 
\lya\ lines detected by HST/GHRS and STIS.  These absorbers appear 
to contain a significant fraction of the low-$z$ baryons and were a major 
discovery of the HST spectrographs.  Using FUSE data, with 
40 m\AA\ ($4\sigma$) \lyb\ detection limits, we have employed the 
equivalent width ratio of \lyb/\lya, and occasionally higher Lyman 
lines, to determine the doppler parameter, $b$, and accurate column 
densities, N$_{\rm HI}$, for moderately saturated lines. We detect 
\lyb\ absorption corresponding to all \lya\ lines with $W_{\lambda} 
\geq 200$ m\AA.  The \lyb/\lya\ ratios yield a preliminary distribution 
function of doppler parameters, with mean 
$\langle b \rangle = 31.4 \pm 7.4$ \kms\ and median 28 \kms, 
comparable to values at redshifts $z = 2.0-2.5$. If thermal, these 
$b$-values correspond to $T_{\rm HI} \approx 50,000$~K,
although the inferred doppler parameters are considerably less than 
the widths derived from \lya\ profile fitting, 
$\langle b/b_{\rm width} \rangle = 0.52$. The typical increase in 
column density over that derived from profile fitting is 
$\Delta$log~N$_{\rm HI} = 0.3$ but ranges up to 1.0 dex. 
Our data suggest that the low-$z$ 
\lya\ absorbers contain sizable non-thermal motions or velocity components 
in the line profile, perhaps arising from cosmological expansion and infall.  


\end{abstract}

\section{INTRODUCTION}

Since the discovery of the high-redshift \lya\ forest over
25 years ago, these abundant absorption features in the spectra
of QSOs have been used as evolutionary probes of the intergalactic
medium (IGM), galactic halos, large-scale structure, and chemical
evolution.  The {\it Hubble Space Telescope} (HST/FOS) Key Project on
QSO absorption systems found that \lya\ absorbers persist to low 
redshift in surprisingly large numbers (Bahcall et al. 1991; 
Morris et al.\ 1991; Jannuzi et al. 1998;  Weymann et al. 1998). 
In this paper, we assume that the \lya\ (and \lyb) lines are
intergalactic.  Richards et al.\ (1999) discuss the possibility
that some C~IV absorption systems could be intrinsic to the AGN
or ejected at relativistic velocities.  

The Colorado group has used HST to conduct
a major survey of \lya\ absorbers 
at low redshift ($z \leq 0.07$) along 15 AGN sightlines, using 
the moderate-resolution (19 \kms) GHRS spectrograph (Stocke et 
al.\ 1995; Shull, Stocke, \& Penton 1996; Penton et al.\ 2000a,b).  
Additional moderate-resolution HST/STIS data along 13 sightlines 
were taken during HST cycle 7.  These observations measured 
\lya\ absorbers down to equivalent widths of 10--20 m\AA\
and determined distributions of the low-$z$ \lya\ absorbers in H~I 
column density for $12.3 < \log N_{\rm HI} < 14.0$ and in line width 
for $15 < b_{\rm width} < 100$ \kms\ (Penton et al.\ 2000b). 
The distribution function, 
$d {\cal N}/dN_{\rm HI} \propto N_{\rm HI}^{-1.80 \pm 0.05}$,
together with photoionization corrections 
suggest (Shull et al.\ 1999a) that the low-$z$ 
\lya\ forest may contain a significant fraction of 
the baryons predicted by nucleosynthesis models of D/H 
(Burles \& Tytler 1998). 

However, a precise baryon census in the low-$z$ IGM requires  
measurements of the true doppler parameter, $b$,
to obtain accurate column densities in saturated \lya\ lines 
with log~N$_{\rm HI} \geq 13.5$.  Measurements of \lyb\ or higher 
Lyman lines are needed to constrain the degree of saturation
through a curve-of-growth (COG) analysis.  
Indeed, recent ORFEUS studies (Hurwitz et al.\ 1998) of \lyb/\lya\ 
ratios in two absorbers toward 3C~373 suggest that $b$ is less than 
the line width determined from \lya\ profile fitting.

With the goal of characterizing the distribution of $b$-values and 
measuring accurate H~I columns, we  conducted a FUSE mini-survey of 
\lyb\ absorbers toward 7 AGN with well-known \lya\ lines.   
The FUSE mission and its capabilities are
described by Moos et al.\ (2000) and Sahnow et al.\ (2000).  
In \S~2 we describe our FUSE \lyb\ observations.  We also
compare simple COG (\lyb/\lya) estimates of $b$ 
with single-component fits to (HST) \lya\ lines
to understand the kinematic structure of the low-$z$ absorbers.  
In \S~3 we present our conclusions and give directions for future 
work on the IGM baryon content, line kinematics, and temperature.

\section{FUSE OBSERVATIONS AND DATA ANALYSIS } 

Our observations were obtained with the FUSE satellite from
1999 September to November during the commissioning phase.
Because the initial FUSE observations were taken with 
relatively long durations (16--49 ksec) we were able to obtain 
reasonable signal-to-noise ratios (S/N $\approx$ 15--20)
in the LiF channels.  The SiC channels had lower S/N 
and were not available for all targets. The spectral 
resolution was generally 20--25 \kms, based on an analysis
of narrow interstellar lines of H$_2$, Ar~I, and Fe~II 
in our H$_2$ mini-survey (Shull et al.\ 2000).    

Our FUSE \lyb\ mini-survey consisted of 7 AGN chosen from targets for 
which the Colorado group had previously obtained HST spectra with 
GHRS/G160M or STIS/E140M.  Our three GHRS targets and full line 
lists are: H1821+643 (Savage, Sembach, \& Lu 1995; Tripp, Lu, 
\& Savage 1998; Penton et al.\ 2000a), ESO~141-G55 (Sembach, 
Savage, \& Hurwitz 1999; Penton et al.\ 2000a), and PKS~2155-304 
(Shull et al.\ 1998; Penton et al.\ 2000a).  Data on our four 
STIS targets (Mrk~876, PG~0804+761, VII~Zw~118, and Ton~S180) 
will be published separately.  

From these 7 targets, we chose 12 strong \lya\ absorbers 
with rest-frame equivalent widths $W_{\lambda} > 200$ m\AA\
and relatively simple line profiles.  
Each of the \lya\ lines had a detectable \lyb\ counterpart, and
in several cases we detected higher Lyman series lines.
Figure 1 shows three examples: the 5512 \kms\ \lyb\ absorber 
toward Ton~S180; the $z = 0.225$ \lyd\ absorber toward 
H1821+643; and the 10,463 \kms\ \lyb\ absorber 
toward ESO~141-G55.  Although PKS~2155-304 contains several strong 
\lya\ lines near $cz = 17,000$ \kms, their line blending is 
sufficiently complex that we did not include them in the survey.
For the 12 chosen lines, we measured equivalent widths
by gaussian profile fitting, using standard FUSE pipeline software.
Assuming a single-component curve of growth, we used the
concordance of the \lyb/\lya\ ratios (or higher Lyman lines)
to derive $b$ and N$_{\rm HI}$.  After minimizing the 
$\chi^2$ of our fits to $N_{\rm HI}$ and $b$, we constructed
$\Delta \chi^2$ contours to define confidence regions
for these inferred quantities.  The error bars on
$N_{\rm HI}$ and $b$ represent single-parameter
68\% confidence intervals for each component and assume that
the true line shape is well represented by a single-component,
doppler-broadened line.  

If \lya\ and \lyb\ are unsaturated, their equivalent widths are
$W_{\lambda}^{Ly\alpha} = (54.5~{\rm m\AA}) N_{13}$ and
$W_{\lambda}^{Ly\beta} = (7.37~{\rm m\AA}) N_{13}$,
where N$_{\rm HI} = (10^{13}~{\rm cm}^{-2}) N_{13}$.
Thus, weak absorbers should have a ratio \lyb/\lya\ = 0.135,
and saturation gradually increases this ratio towards 1.
Our typical FUSE \lyb\ detection limit of 40 m\AA\ ($4 \sigma$)
corresponds to log~N$_{\rm HI} \geq 13.73$.   Our success in detecting 
\lyb\ for all \lya\ lines with $W_{\lambda} > 200$ m\AA\ is 
consistent with the relative line strengths.

Table 1 lists the 12 lines in our survey, together with \lya\ 
velocities ($cz$) and line widths ($b_{\rm width}$) derived from 
instrumentally corrected gaussian fits to the GHRS or STIS \lya\ 
profiles. Here, $b_{\rm width} = 2^{1/2} \sigma_{\rm gauss}$,  
and $b$ and N$_{\rm HI}$ are derived from the COG.  
Figures 2 and 3 show COG concordance plots for two absorbers in 
Fig.\ 1.  Not including the broad, blended line at 16,203 \kms\ 
toward PKS~2155-304, the FUSE distribution has mean 
$\langle b \rangle = 31.4 \pm 7.4$ km~s$^{-1}$
and median 28~km~s$^{-1}$, comparable to the values, 
$\langle b \rangle = 27.5 \pm 1.3$ \kms\ and median
26.4 \kms, measured in the $z =$ 2.0--2.5 \lya\ 
forest (Rauch et al.\ 1993). If thermal, these $b$-values 
correspond to $T_{\rm HI} = m_H b^2/2k \approx 50,000$~K, a temperature 
higher than values predicted from models of low-metallicity \lya\ clouds 
(Donahue \& Shull 1991), heated and photoionized by the
AGN metagalactic background.  These models predict temperatures and 
doppler parameters, conveniently approximated by 
$T = (24,300~K)(U/0.01)^{0.152}$ and $b = (20~{\rm km~s}^{-1})
(U/0.01)^{0.076}$.  The doppler parameter scales 
weakly with the photoionization parameter 
$U = n_{\gamma}/n_H \approx (0.005) J_{-23} 
(10^{-5}~{\rm cm}^{-3}/n_H)$. Here, $U$ is the ratio of ionizing 
photons to hydrogen nuclei for a specific intensity 
$J_0 = (10^{-23}$ ergs s$^{-1}$ cm$^{-2}$ sr$^{-1}$ Hz$^{-1}) J_{-23}$.

The doppler parameters inferred from \lyb/\lya\ are considerably less 
than widths derived from \lya\ profile-fitting.  The statistical 
average, $\langle b/b_{\rm width} \rangle = 0.52$, suggests that the 
\lya\ and \lyb\ line profiles are more complex than single-component,
doppler-broadened gaussians. The widths of individual 
components must be less than the COG-inferred $b$-values,
a general principle exemplified by the special case where
$N$ identical, well-separated gaussians behave as a
single one with a dispersion equal to $Nb$. 
For each ensemble of components, the total column density derived
from \lya\ and \lyb\ might be larger than the true value,
but not by a large factor (Jenkins 1986).  

The \lya\ absorbers could be broadened by cosmological expansion, 
$\Delta v = (19.5~{\rm km~s}^{-1})(\Delta r/300~{\rm kpc})h_{65}$, 
across spatially extended H~I absorbers, as seen in  
numerical simulations (Weinberg, Katz, \& Hernquist 1998).  The 
profiles could also be blends of velocity components arising from 
clumps of gas falling into dark-matter potential wells.
If these clumps are sufficiently massive to be gravitationally bound, 
the velocity components may represent small-scale power at 
$\Delta v < 100$ \kms\ in the cosmological spectrum of density 
fluctuations.  We have seen spectral evidence for such velocity components in
the low-$z$ \lya\ absorbers studied by HST/GHRS and STIS
at 19 \kms\ resolution and in their two-point correlation function, 
$\xi (\Delta v)$ (Penton et al.\ 2000b,c).
We have also verified, through optical and 21-cm imaging, that
some \lya\ absorbers arise within small groups (van Gorkom et al.\
1996; Shull et al.\ 1998).  
  
With better statistics on the \lya\ and \lyb\ line widths
and doppler parameters, we should be able to constrain the kinematics 
of the absorber profiles. A critical issue is whether cosmological
expansion or velocity components are the dominant contributor to
line broadening.  Recent numerical simulations of the 
low-$z$ \lya\ forest (Dav\'e et al.\ 1999) suggest that both effects 
are important.

\section{CONCLUSIONS AND DISCUSSION}

First and foremost, FUSE has identified the low-redshift 
\lyb\ forest.  Lines between 1216~\AA\ and (1216~\AA)$(1+z_{\rm em}$) 
with no clear identification are often labeled as \lya.  The FUSE 
detection of their \lyb\ counterparts makes these identifications 
conclusive. We have used the COG concordance of Ly$\beta$/Ly$\alpha$
equivalent widths, and occasionally higher Lyman lines,
to derive reliable values of $b$ and N$_{\rm HI}$
for the stronger, saturated lines ($W_{\lambda} > 200$ m\AA).
Our major findings are:
(1) The doppler parameters from single-component COG fits to
\lyb/\lya\ equivalent widths are considerably
less than those derived from line profile fitting,
with  $\langle b/b_{\rm width} \rangle = 0.52$;
(2) We find $\langle b \rangle = 31.4 \pm 7.4$ \kms\ 
(median 28 \kms), similar to values at redshifts $z =$ 2.0--2.5;
(3) Although these $b$-values correspond to $T_{\rm HI} \approx 
50,000$~K, the low-$z$ absorbers may 
contain non-thermal motions or line broadening 
from cosmological expansion and infall.

Over its lifetime, FUSE will observe many AGN
sightlines for the O~VI, D/H, and AGN projects.
This will produce a large survey of Ly$\beta$ absorbers
at $z < 0.155$ that we can use to characterize the
distributions in $b$ and N$_{\rm HI}$ in the
low-redshift \lyb\ forest.  The distribution function, $f(b)$,
can be used to derive the IGM temperature distribution
and infer its equation of state (Schaye et al.\ 1999;
Ricotti, Gnedin, \& Shull 2000). 
With high-S/N data and a flux-limited sample,
we can search for the hot baryons predicted by
cosmological simulations (Cen \& Ostriker 1999a).
These absorbers should appear in the high-$b$ tail of the
distribution as broad, shallow absorbers.

Because \lya\ lines with  $W_{\lambda} > 130$ m\AA\ 
(log N$_{\rm HI} > 13.5$)  appear to be saturated 
(Penton et al.\ 2000b), HST alone cannot provide accurate column 
densities for the strong \lya\ absorbers that probably dominate 
the baryon content and opacity of the low-$z$ IGM.  For lines 
in which $b$ is well determined, log~$N_{\rm HI}$ 
typically increases by 0.3 dex, and up to 1 dex,  
compared to \lya\ profile fitting.  This increase, which is 
greatest in the most saturated \lya\ lines, means that an even larger 
fraction of baryons may be found in the low-$z$ \lya\ forest.  

A full \lyb\ survey will also provide a sample of high-$N_{\rm HI}$ 
absorbers  which can be used to estimate the level of metallicity in 
the low-$z$ IGM  (Shull et al.\ 1998; Cen \& Ostriker 1999b). 
As an illustration, we constructed simple 
photoionization models using CLOUDY (Ferland 1996) that assume 
a specific ionizing intensity at 13.6 eV of
$J_\nu = 10^{-23}$ ergs cm$^{-2}$ s$^{-1}$ sr$^{-1}$ Hz$^{-1}$ 
(Shull et al.\ 1999b) with $J_\nu \propto \nu^{-1.8}$ and 
column density $\log N_{\rm HI} = 15$.  For total hydrogen 
densities (cm$^{-3}$) of $\log n_H = (-3,-4,-5)$, a 20 m\AA\ measurement 
of C~III $\lambda$977 implies $Z/Z_{\odot} = (0.3, 0.15, 0.01)$. For 
$\log n_H = (-5, -6)$, a 20 m\AA\ measurement of O~VI $\lambda$1032
implies $Z/Z_{\odot} = (0.015, 0.006)$.
Observations of C~III $\lambda977$ will be especially
useful, because they can be compared with
C~II and/or C~IV to obtain reliable ionization corrections.


\acknowledgements 

This work is based on FUSE Team data obtained for the Guaranteed Time
Team by the NASA-CNES-CSA FUSE mission operated by Johns Hopkins University.
Financial support to U.~S. participants has been provided by NASA
contract NAS5-32985. The Colorado group also acknowledges support from
astrophysical theory grants from NASA (NAG5-7262) and NSF (AST96-17073),
the HST/COS project (NAS5-98043), and STScI grant GO-0653.01-95A
which supported the \lya\ data analysis. 



\begin{deluxetable}{lcccccccc}
\footnotesize
\tablecaption{FUSE Ly$\beta$ Lines\tablenotemark{1} \label{tbl:Lyb} }
\tablehead{
\colhead{AGN} &
\colhead{$t_{\rm exp}$} &
\colhead{\lya\ Velocity } &
\colhead{$W_\lambda$(Ly$\alpha$) } &
\colhead{$W_\lambda$(Ly$\beta$) }  &
\colhead{$\log N_{HI}^{Ly\alpha}$ } &
\colhead{$b_{\rm width}^{Ly\alpha}$    } &
\colhead{$\log N_{HI}$ } &
\colhead{$b$} \nl
\colhead{}    &
\colhead{ (ksec) } &
\colhead{(km~s$^{-1}$) } &
\colhead{ (m\AA)}    &
\colhead{ (m\AA)} &
\colhead{ (cm$^{-2}) $} &
\colhead{ (km~s$^{-1}$)} &
\colhead{ (cm$^{-2}) $} &
\colhead{ (km~s$^{-1}$)}
}
\startdata
ESO 141-G55   & 35.8 & 10,463\tablenotemark{2} & 354$\pm$11 & 164$\pm40$ & 
   14.06$\pm 0.03$ & 49$\pm$3 & 14.57$^{+0.25}_{-0.23}$ & 27$^{+5}_{-3}$  \nl
H1821+643     & 48.8 & 7342 & 298$\pm$20  & 90$\pm$30 & 13.93$^{+0.05}_{-0.04}$ &
   49$\pm$3   & 14.23$^{+0.19}_{-0.18}$ & 27$^{+8}_{-4}$ \nl
         &     & 63,933 & 483$\pm$18 & 155$\pm$18 & 14.31$^{+0.08}_{-0.06}$&
   49$\pm$2   & 14.43$^{+0.07}_{-0.06}$ & 43$^{+3}_{-3}$ \nl
         &    & 67,420\tablenotemark{3} & 739$\pm$22 & 511$\pm$21& 
          14.41$\pm 0.04$& 87$\pm$2 & 15.46$^{+0.06}_{-0.05}$ & 45$^{+2}_{-1}$ \nl
Mrk 876       & 45.9   & 958 & 324$\pm$52  &170$\pm$40 & 13.89$\pm 0.11$  &
     72$\pm$8 & 14.42$^{+0.09}_{-0.13}$ & 26$^{+6}_{-5}$  \nl
PG~0804+761   & 39.6 & 5565 & 329$\pm$18 & 90$\pm$30 & 13.94$^{+0.03}_{-0.04}$ &
       57$\pm$2 & 14.18$^{+0.17}_{-0.24}$ & 32$^{+26}_{-5}$  \nl
VII~Zw~118    & 26.0 & 2463 & 355$\pm$35 & 110$\pm$50 & 14.01$^{+0.13}_{-0.10}$ &
       54$\pm$8 & 14.28$^{+0.27}_{-0.34}$ & 32$^{+49}_{-7}$ \nl
PKS~2155-304  & 37.1 & 5119  & 218$\pm$20 & 40$\pm$20 & 13.68$^{+0.04}_{-0.05}$ &
   82$\pm$5 & 13.76$^{+0.22}_{-0.13}$ & 41$^{+\infty}_{-20}$ \nl
         &    & 16,203 & 346$\pm$23  & 40$\pm$20 & 13.96$\pm 0.05$ & 60$\pm$3  &
       13.88$^{+0.09}_{-0.03}$ & 98$^{+\infty}_{-44}$  \nl
TON~S180      & 16.5 & 5512  &271$\pm$16  &71$\pm$15 & 13.82$\pm0.04$  &
      58$\pm$2 & 14.05$^{+0.12}_{-0.12}$ & 28$^{+9}_{-4}$  \nl
        &   & 7025 &236$\pm$17  & 65$\pm$40 &  13.72$\pm 0.04$ & 67$\pm$3 &
       14.03$^{+0.32}_{-0.35}$ & 23$^{+\infty}_{-5}$  \nl
       &    & 13,688 & 229$\pm$17  & 55$\pm$15 &  13.72$\pm 0.04$ & 63$\pm$3 &
      13.94$^{+0.13}_{-0.18}$ & 25$^{+22}_{-5}$  \nl
\enddata
\tablenotetext{1}{Columns, left to right: (1) AGN sightline; (2) FUSE exposure 
  time; (3) recession velocity ($cz$) relative to LSR; (4, 5) rest-frame \lya\ 
  and \lyb\ equivalent widths with 1$\sigma$ uncertainties; (6) H~I column 
  density from fit of single-component, doppler-broadened and 
  instrumentally broadened \lya\ profile; (7) instrumentally corrected
  \lya\ line width and 1$\sigma$ uncertainty; (8, 9) H~I column 
  density and doppler parameter, based on \lyb/\lya\ equivalent width
  ratio (exceptions below).  Uncertainties represent 68\% confidence interval.  
  FUSE observations of higher Lyman series lines were also used in analysis 
  of the following:
     H1821+643, $cz = 7344$ kms$^{-1}$, $W_\lambda$(\lyg) = $57\pm30$ m\AA;
     H1821+643, $cz = 63933$ kms$^{-1}$, $W_\lambda$(\lyg) = $62\pm25$ m\AA;
     H1821+643, $cz = 67420$ kms$^{-1}$, $W_\lambda$(\lyd) = $212\pm25$ m\AA,
                 $W_\lambda$(\lye) = $139\pm 25$ m\AA;
     Mrk 876, $cz = 958$ kms$^{-1}$, $W_\lambda$(\lyg) = $40\pm20$ m\AA.
}
\tablenotetext{2} {Associated \lya\ absorber; \lyg\ indicates 
    two velocity components, separated by 70 \kms}
\tablenotetext{3} {\lyd\ (Fig. 2) indicates two velocity components, 
    separated by 70 \kms}
\end{deluxetable}

\newpage

\begin{figure*}[t]
  \centerline{\epsfxsize=0.7\hsize{\epsfbox{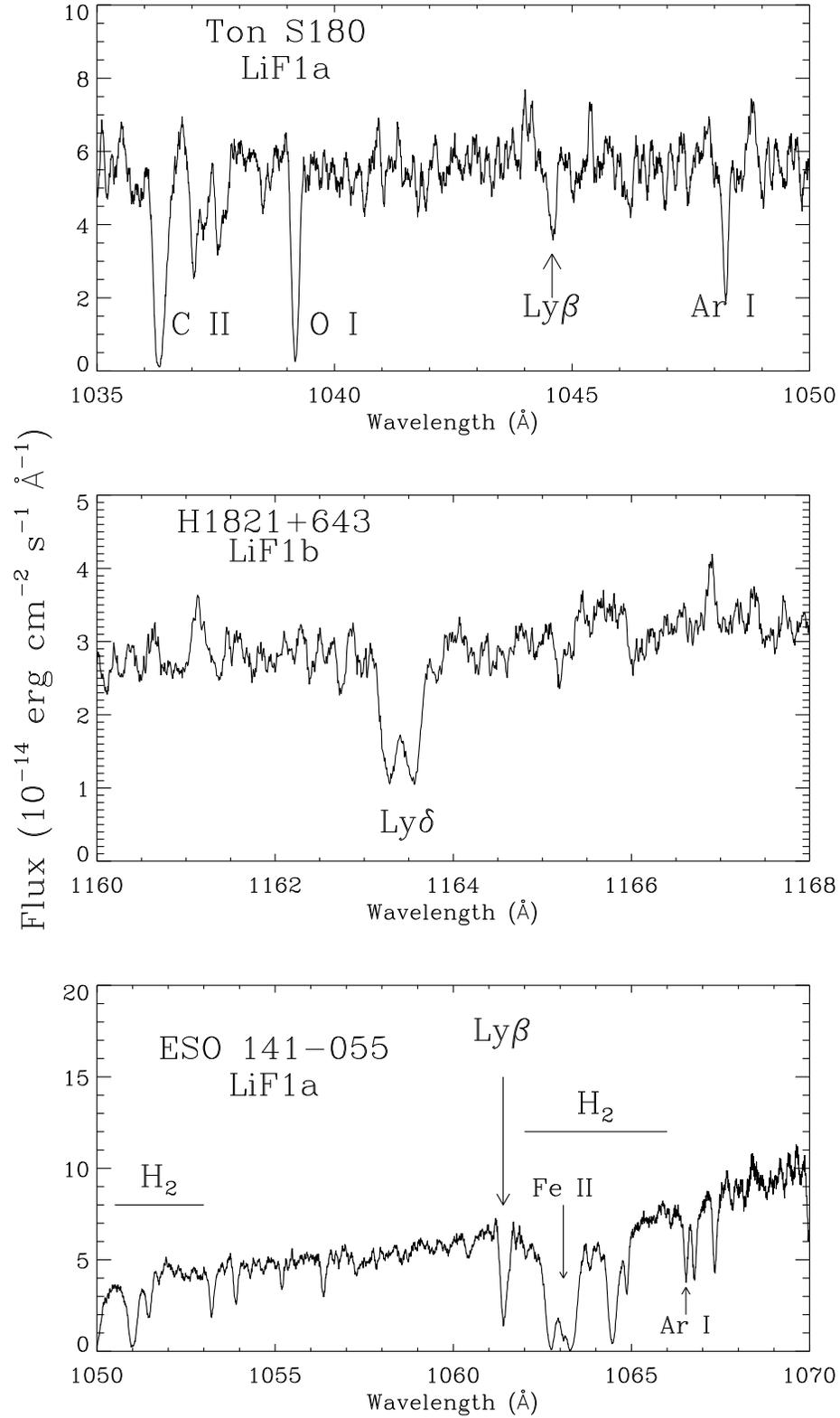}}}
  \figcaption{FUSE spectra of three absorbers toward targets
   Ton~S180 (\lyb\ in 5512 \kms\ absorber),
   H1821+643 (\lyd\ in $z = 0.225$ absorber), and
   ESO~141-G55 (\lyb\ in 10,463 \kms\ absorber).
   The H1821+643 \lyd\ absorber shows two components
   separated by 70 \kms.  }
\label{fig1}
\end{figure*}

\begin{figure*}[t]
\centerline{\epsfxsize=\hsize{\epsfbox{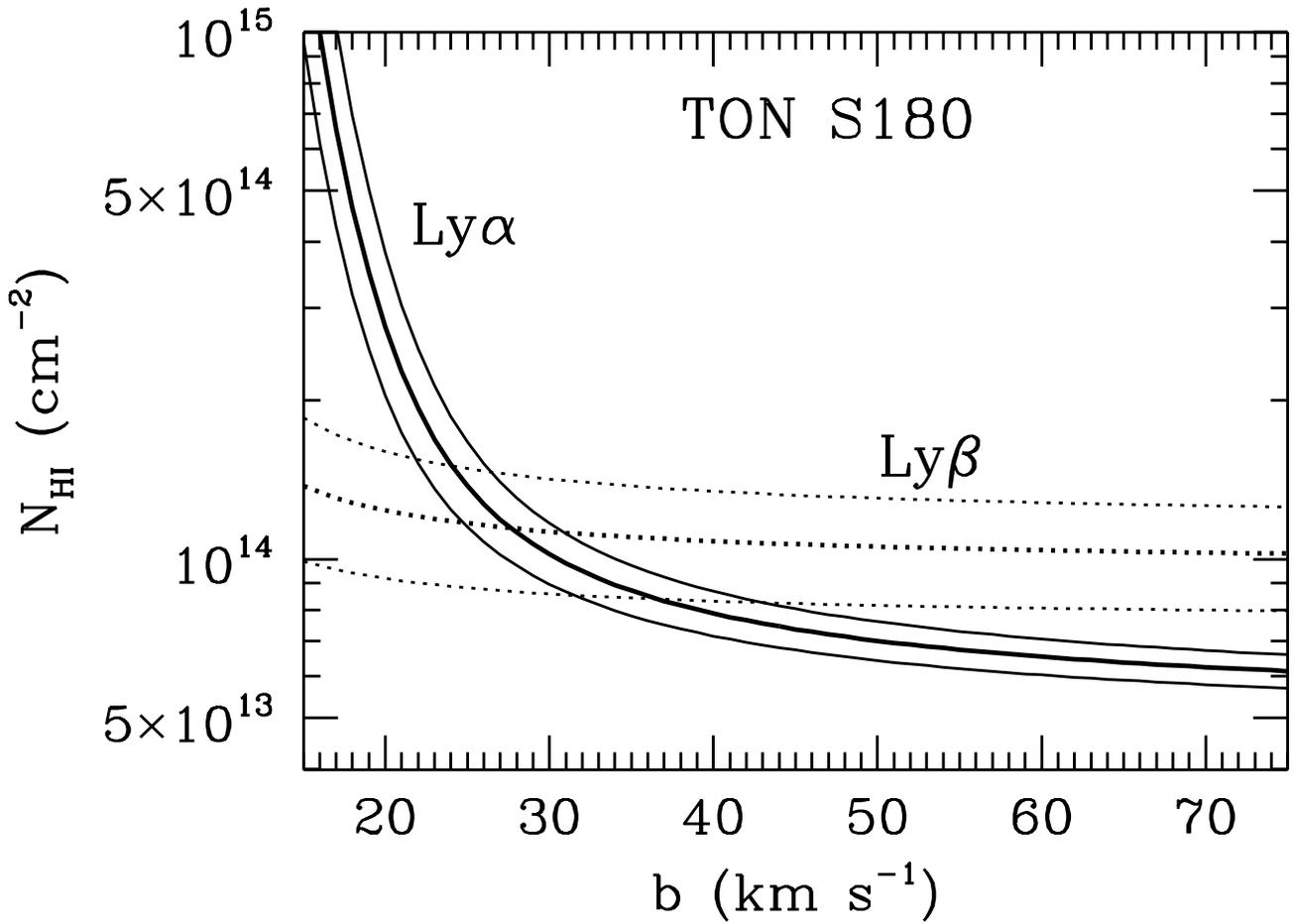}}}
\figcaption{COG concordance plot in $b$ and N$_{\rm HI}$ for \lya\
   ($271 \pm 16$ m\AA) and \lyb\ ($71 \pm 15$ m\AA) at
   $cz = 5512$ \kms\ toward Ton S180.  Three curves for each line
   show single-component, doppler-broadened COG fits to equivalent
   widths ($\pm 1 \sigma$ errors). }
\label{fig2}
\end{figure*}

\begin{figure*}[t]
\centerline{\epsfxsize=\hsize{\epsfbox{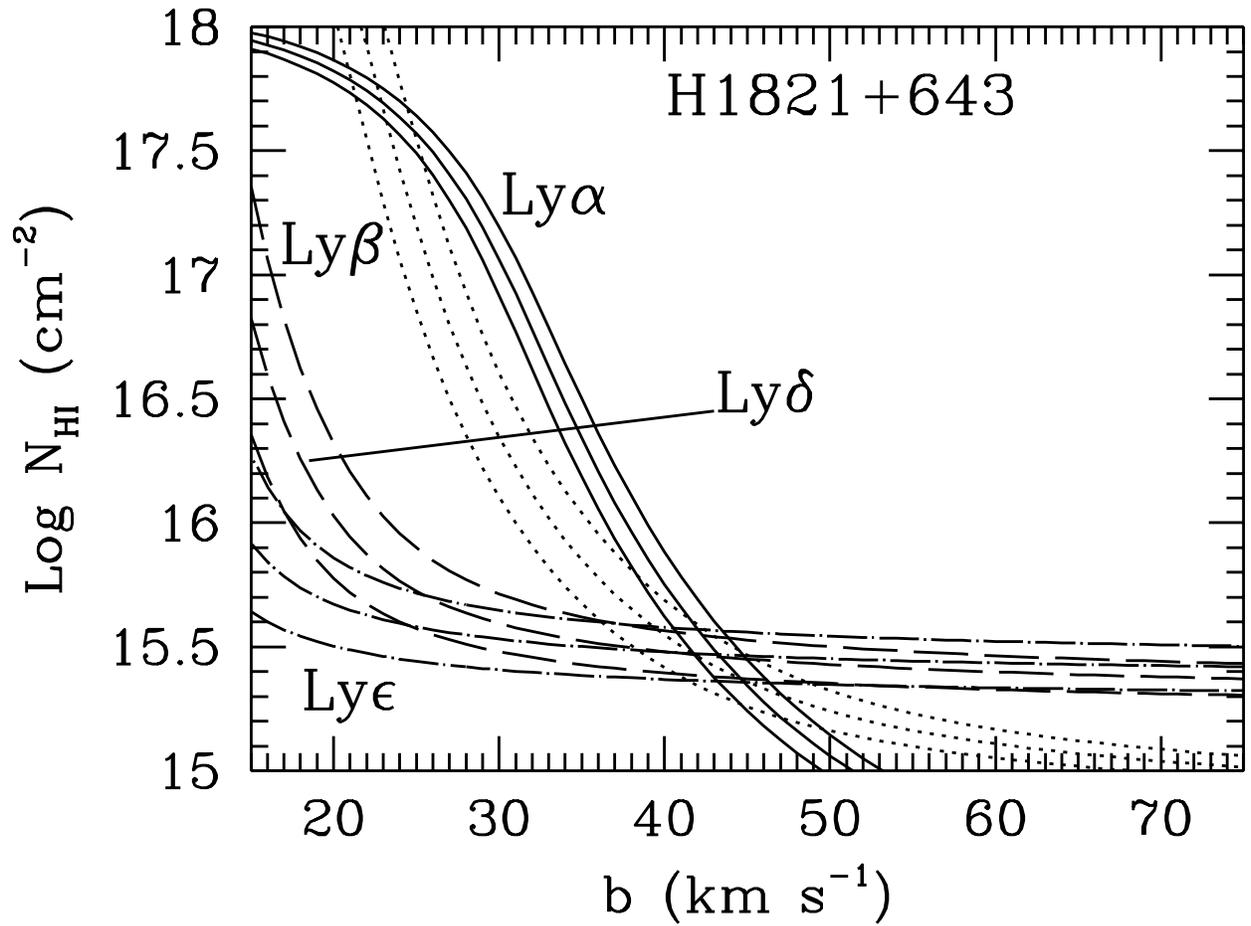}}}
\figcaption{COG concordance plot (see Fig.\ 2) for absorber at $z = 0.225$ toward
  H1821+643.  The curves of growth are based on rest-frame equivalent
   widths for \lya\ ($739 \pm 22$ m\AA) and \lyb\ ($511 \pm 21$ m\AA)
   from HST/GHRS/G160M and for \lyd\ ($212 \pm 25$ m\AA) and \lye\
   ($139 \pm 25$ m\AA) from FUSE. }
\label{fig3}
\end{figure*}

\end{document}